\documentclass[default,iicol]{sn-jnl}% Default with double column layout

\usepackage{graphicx}
\usepackage{subfigure}
\usepackage{float}
\usepackage{amsmath,amsfonts}
\jyear{2021}%

\theoremstyle{thmstyleone}%
\theoremstyle{thmstyletwo}%

\theoremstyle{thmstylethree}%
\begin{document}

\title[Article Title]{Can LSH (Locality-Sensitive Hashing) Be Replaced by Neural Network?}


\author[1,3]{\fnm{Renyang} \sur{Liu}}\email{ryliu@mail.ynu.edu.cn}
\equalcont{This paper was accepted by Soft Computing.}

\author[4]{\fnm{Jun} \sur{Zhao}}\email{jhinzhao@didiglobal.com}

\author[2,3]{\fnm{Xing} \sur{Chu}}\email{chx@ynu.edu.cn}

\author[2,3]{\fnm{Yu} \sur{Liang}}\email{yuliang@ynu.edu.cn}

\author[2,3]{\fnm{Wei} \sur{Zhou}}\email{zwei@ynu.edu.cn}

\author*[2,3]{\fnm{Jing} \sur{He}}\email{hejing@ynu.edu.cn}

\affil*[1]{\orgdiv{School of Information Science and Engineering}, \orgname{Yunnan University}, \orgaddress{\street{Wujiaying}, \city{Kunming}, \postcode{650500}, \state{Yunnan}, \country{China}}}

\affil[2]{\orgdiv{School of Software}, \orgname{Yunnan University}, \orgaddress{\street{Wujiaying}, \city{Kunming}, \postcode{650500}, \state{Yunnan}, \country{China}}}

\affil[3]{\orgdiv{Engineering Research Center of Cyberspace}}

\affil[4]{\orgname{Didi Chuxing}, \orgaddress{\street{Dongbeiwang},  \postcode{100000}, \state{Beijing}, \country{China}}}

\abstract{With the rapid development of GPU (Graphics Processing Unit) technologies and neural networks, we can explore more appropriate data structures and algorithms. Recent progress shows that neural networks can partly replace traditional data structures. In this paper, we proposed a novel DNN (Deep Neural Network)-based learned locality-sensitive hashing, called LLSH, to efficiently and flexibly map high-dimensional data to low-dimensional space. LLSH replaces the traditional LSH (Locality-sensitive Hashing) function families with parallel multi-layer neural networks, which reduces the time and memory consumption and guarantees query accuracy simultaneously. The proposed LLSH demonstrate the feasibility of replacing the hash index with learning-based neural networks and open a new door for developers to design and configure data organization more accurately to improve information-searching performance. Extensive experiments on different types of datasets show the superiority of the proposed method in query accuracy, time consumption, and memory usage.}


\maketitle

\section{Introduction}
\label{intro}
Given a set of data points and a query, searching for the nearest data point in a given database is the fundamental problem of NN (Nearest Neighbor) search \cite{DBLP:journals/paa/Moraleda08,DBLP:journals/soco/BhaskarK20,DBLP:journals/soco/BeheraK21}, which is widely used in information retrieval, data mining, multimedia, and scientific databases. Suppose there is a query point \emph{q} and dataset \emph{D}, the NN problem is to find an item $q_{1}$ from $D$ on the condition that the distance between $q_{1}$ and $q$ is closest. One extension of NN is kNN that find top-\emph{k} closest items to the query data $\{q_{1},...,q_{k}\}$ from $D$. The traditional kNN algorithm is mainly based on spatial division, which is most widely used in the tree algorithms, such as KD-tree \cite{DBLP:journals/tse/Bentley79}, R-tree \cite{DBLP:conf/sigmod/Guttman84}, Ball-tree \cite{1978A}. Although the query accuracy of the tree-based approach is high, they require a huge amount of memory, sometimes even exceeding the data itself. Besides, the performance of tree-based indexing methods will be significantly faked when handling high-dimensional data \cite{DBLP:reference/cg/2004,DBLP:journals/soco/NguyenDMG21}, which is named ``curse of dimensionality" \cite{2014Fundamentals}. In addition, with the development of the current business systems, the data dimensions are increasing, achieving from thousands to millions. It puts a high demand on finding a new way to deal with kNN efficiently because the traditional indexing methods are challenging to handle the high dimensional data.

One feasible way is to transform the NN and kNN problems into ANN (Approximate Nearest Neighbors) and kANN problems to cope with the growing data dimension. In the ANN search, the index method only needs to return the approximate nearest objects $\{q_{1},...,q_{k}\}$ rather than find the actual nearest one. In this way, the query efficiency can be significantly improved. The ANNs have a lot of advantages to solving the search tasks in scenarios which not require high precision to reduce time and memory consumption. Among them, the LSH \cite{DBLP:conf/stoc/IndykM98}, which basic principle is that the two adjacent data points in the original data space can be hashed into the same bucket by the same mapping or projection transformation rule is the most popular one. And it is widely used in various searching fields, including and not limited to text, audio, image, video, gene, et al., due to its unusual nature of locality sensitivity and the superiority to KD-tree \cite{DBLP:journals/tse/Bentley79} and other methods in high-dimensional searching.

Traditional LSH, however, is applied to CPU, parallel computing and distributed applications, which greatly limits its potential in the face of high dimensional data. Moreover, due to the rapid development of hardware, such as GPU/TPU (Tensor Processing Unit), the high cost of performing neural networks may be negligible in the near future. Therefore, inspired by the pioneering work \cite{DBLP:conf/sigmod/KraskaBCDP18} in developing a learned index to explore how neural networks can enhance or even replace traditional index structures. In this paper, we design a novel neural networks-based framework, called LLSH, to boot the E2LSH (Exact Euclidean Locality Sensitive Hashing) \cite{DBLP:conf/compgeom/DatarIIM04} in the task of massive data retrieval. The LLSH creatively proposed to replace the hash functions in E2LSH with a simple neural network to improve the search efficiency of the hash indexing. Extensive experiments illustrated the proposed framework's feasibility and superiority in query accuracy, time, and memory consumption. The main contributions of this paper are reflected as follows:

$\triangleright$ We propose a novel DNN-based learned locally-sensitive hashing, called LLSH, which can be applied to the kNN problem of high-dimensional data and avoid "dimensional curses." To the best of our knowledge, it is the first work to use neural networks instead of hash function families. Each neural network is independent and computes parallelly to fully utilize the hardware's advantages and reduce the false-positive and false-negative rates.

$\triangleright$ We design the framework of LLSH in detail and apply it to replace the traditional E2LSH with two different strategies. The basic one trains the neural network layer supervised by the E2LSH outputs, while the ensemble one takes a forward step to fully utilize the idea of ensemble learning to integrate the outputs of multiple NN algorithms to improve the performance.

$\triangleright$ We conduct extensive experiments, which include feasibility verification, time and memory consumption, and query accuracy, on eight datasets with different data types and distributions. The empirical results show the viability of the proposed LLSH framework and its superiority in reducing time and memory usage and improving query accuracy.

The rest of the paper is organized as follows. We briefly review the methods relating to data structure and machine learning in Sec. \ref{Sec:related}. In Sec. \ref{preliminary}, we provide the preliminaries of LSH and E2LSH. Sec. \ref{method} discusses the details of the proposed LLSH. The experimental results are shown and analyzed in Sec. \ref{experiment}. Finally, the paper is concluded in Sec. \ref{conclusion}.

\section{Related Works}
\label{Sec:related}
Our work is based on a wide range of previous excellent research. In the following, we intend to summarize several essential interactions between data structure and machine learning.

LSH is a hashing algorithm that was first proposed by Indyk in 1998. In general, the hash algorithm is a way to reduce conflicts, and it can facilitate quick additions and deletions, but LSH is not. LSH, which uses the hash conflict to speed up the retrieval effect, is mainly applied to the fast approximate search of high-dimensional mass data. The approximate search is a comparison of distances or similarities between data points. According to the different methods of similarity calculation, LSH can be divided into several categories, including Simhash \cite{DBLP:conf/www/MankuJS07}, E2LSH \cite{DBLP:conf/compgeom/DatarIIM04}, C2LSH \cite{DBLP:conf/sigmod/GanFFN12}, Kernel LSH \cite{DBLP:conf/iccv/KulisG09}, LSB-forest \cite{DBLP:conf/sigmod/TaoYSK09}, QALSH \cite{DBLP:journals/pvldb/HuangFZFN15} etc.

LSH families have many branches and are widely used in various applications. For example, Simhash maps the original text content to a digital hash signature, where the two similar texts correspond to the same digital signature. So, the similarity of the two documents can be measured by the Hamming distance between the Simhash value. E2LSH is a randomized implementation method of LSH in Euclidean space. The basic principle of E2LSH is to use the position-sensitive function based on p-stable distribution to map the high-dimensional data and keep the two neighbor points in the original space still closest to each other after the mapping operation. LSB-forest builds multiple trees to adjust to the NN search. Sun et al. devised SRS \cite{DBLP:journals/pvldb/SunWQZL14} with a small index footprint so that the entire index structure can fit in lesser memory. Recently, a new LSH scheme named QALSH (Query-aware data-dependent LSH) has been proposed to improve search accuracy by deciding the bucket boundaries after the query arrives at its position. 

\begin{figure*}[ht]
	\includegraphics[width=\linewidth]{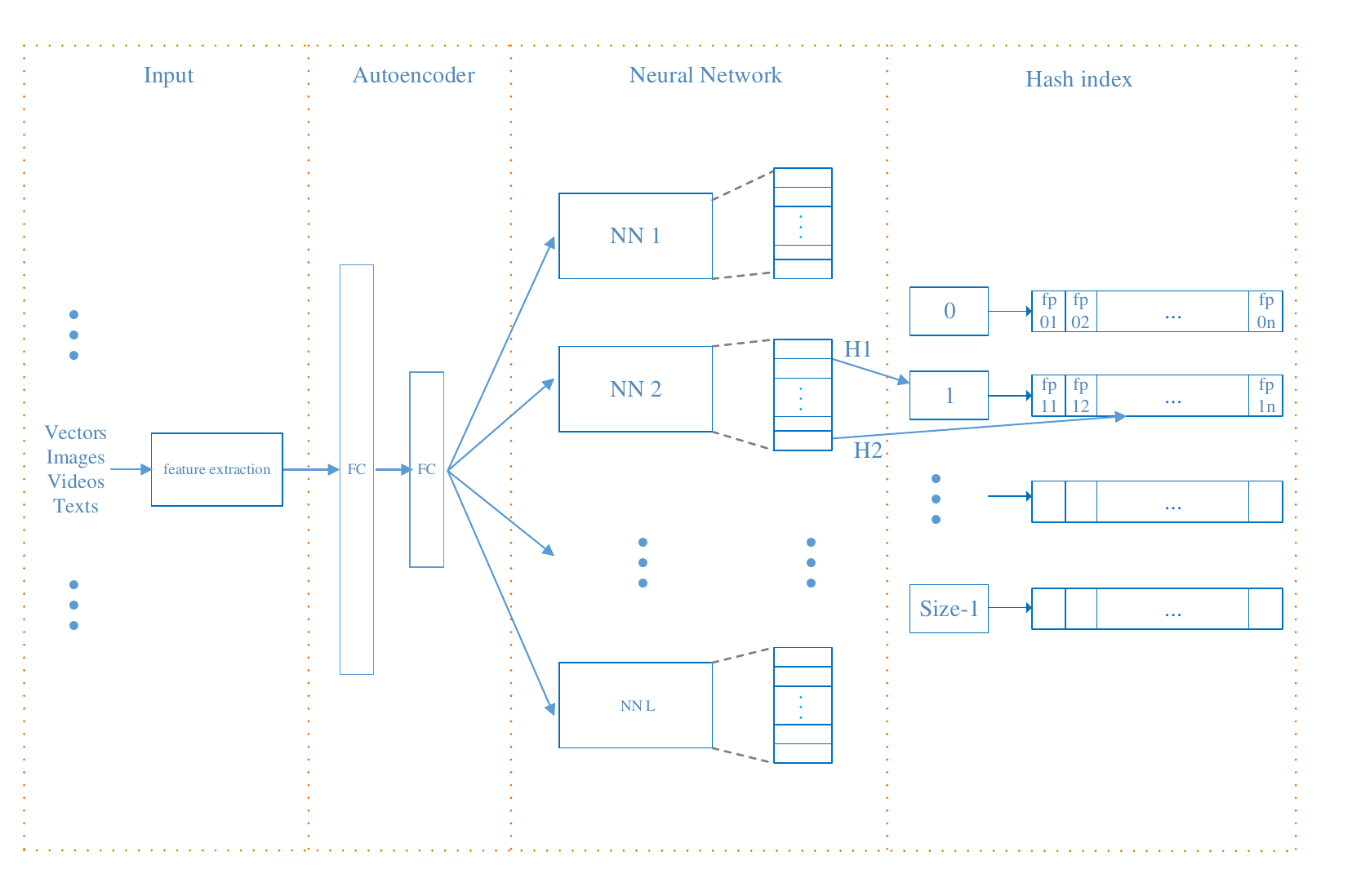}
	\caption{The framework of DNN-based learned index.}
	\label{fig:1}
\end{figure*}

However, with the development of AI (Artificial Intelligence) and the explosion of data complexity, machine learning has become a powerful technique for solving computer optimization problems, which require new methods to compute more efficiently and intelligently. Recently, researchers have begun employing machine learning to optimize indexes and hash functions. There is various research on emulating locality-sensitive hash functions to build the new ANN indexes, ranging from supervised \cite{DBLP:conf/cvpr/LiuWJJC12,DBLP:conf/nips/0002FS12,DBLP:conf/cvpr/TorralbaFW08,DBLP:journals/ijon/ChenSYXS17} to unsupervised \cite{DBLP:conf/icml/LiuWKC11,DBLP:journals/pami/GongLGP13,DBLP:conf/nips/KongL12,DBLP:conf/nips/GongKVL12,DBLP:journals/ijon/JinYSZ19}. These kinds of methods incorporate data-driven learning methods in developing advanced hash functions. The principle of these works is learning to a hash, which means learning the information of data distributions or class labels to guide the design of the new learning-based hash function. However, the hash function's basic construction is still unchanged. Although, there are some methods, like \cite{DBLP:conf/cvpr/LinYHC15,DBLP:conf/aaai/XiaPLLY14}, using the neural network to replace a hash function and using the image as the hash label to pursue the good search performance in image retrieval, but these limits the scope of the hash method and cannot be used to construct fundamental data structures directly. 

As far as we know, paper \cite{DBLP:conf/sigmod/KraskaBCDP18} is the pioneering work in developing a learning index that explores how neural networks can enhance and even replace traditional index structures. It provides a learned index based on a neural network to replace the B-tree index and further discusses the difference between learning hash mapping and traditional hash mapping index. Moreover, Our previous work also provides an unsupervised learned index named PAVO \cite{DBLP:journals/access/XiangZCCLZ19}. Therefore, we are well motivated by these works to propose a novel neural network-based learned hash index framework that can utilize new techniques, like a deep neural network, and new hardware, like high-performance GPU, to construct a novel learning-based hash method for massive magnitude and dimensional data retrieve. 

\begin{figure*}[ht]
	\includegraphics[width=\linewidth]{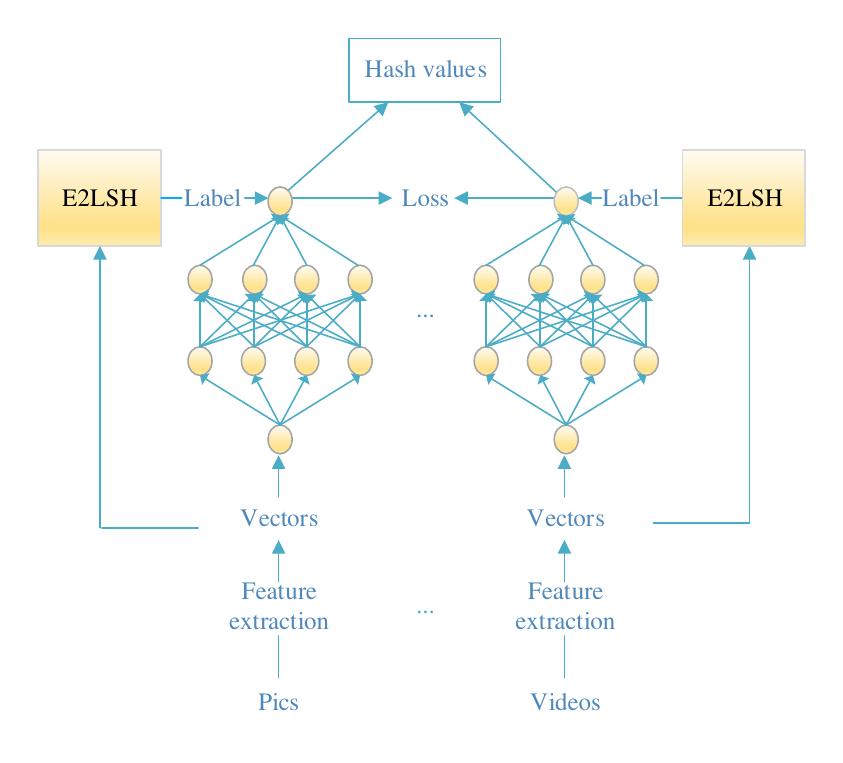}
	\caption{The supervised strategy in neural network stage.}
	\label{fig:2}
\end{figure*}

\section{Preliminary}
\label{preliminary}
LSH is a fast nearest neighbor search algorithm for massive high-dimensional data. We call such a family of hash functions $H={h:S \rightarrow U}$ as $(r_{1},r_{2},p_{1},p_{2})$ sensitive if the function $h$ in any $H$ satisfies the following two conditions:
$${if\,d(O_{1},O_{2})<r_{1}\,then\,Pr[h(O_{1})=h(O_{2})]\geq{p_{1}}},$$
$${if\,d(O_{1},O_{2})>r_{2}\,then\,Pr[h(O_{1})=h(O_{2})]\leq{p_{2}}}.$$

Among them, $O_1, O_2 \in S$, denote two data objects with multi-dimensional attributes, $d(O_{1}, O_{2})$ is a metric function that represents the degree to which two objects are different. And the threshold $(r_{1},r_{2},p_{1},p_{2})$ satisfies the condition: $r_{1}<r_{2}$ and $p_{1}>p_{2}$. It means that two high-dimensional data are mapped to the same hash values when they are similar enough.

The LSH can be divided into different types according to the different similarity calculation methods. One of the most widely used is the p-stable hash, also called E2LSH, which uses a Euclidean distance to measure data similarity. The p-stable distribution refers to a type of distribution defined as follows.

For any $n$ real numbers $v_{1},v_{2},...,v_{n}$ and $n$ random variables $d_{1},d_{2},...,d_{n}$ subject to the distribution \emph{D}, there is a $p\geq 0$ that makes $\sum_i{v_id_i}$ and $(\sum_i{ v_{i} ^{p}})^{1/p}$ have the same distribution ($d$ is a random variable in the p-stable distribution). For E2LSH, the \emph{p} of the p-stable distribution is limited to $ 0<p\leq{2}$ and defined as follows:

$\triangleright$ 1-stable: Cauchy Distribution
\begin{equation}
	c(x)=\frac{1}{\pi}\frac{1}{1+x^2};
\end{equation}

$\triangleright$ 2-stable: Gaussian Distribution
\begin{equation}
	g(x)=\frac{1}{\sqrt{2\pi}}e^{-x^2/2}.
\end{equation}

The family of hash functions are proposed as follows \cite{DBLP:conf/compgeom/DatarIIM04}:
\begin{equation}
	h_{a,b}(v)=\lfloor{\frac{av+b}{r}\rfloor,}
\end{equation}
where $a$ is a vector that conforms to the p-stable, and the dimension is the same as $v$, $b \in (0,r)$ is a random number, $r$ is the length of a straight line segment, the establishment of hash function family is based on the differences of $a$ and $b$.

So, if two points $v_1$ and $v_2$ are supposed to be mapped into the same hash value, they must satisfy $av_1+b$ and $av_2+b$ are mapped to the same line segment.

Let $f_p(t)$ denote the probability density function of the absolute value of the p-stable distribution. For two vector $v_1,v_2$, make $ c= \left \| v_1-v_2 \right \|_p $, the collision probability in E2LSH is calculated as follows:

\begin{equation}
	p(c)  =P_{a,b}[h_{a,b}(v_{1})=h_{a,b}(v_{2})]          
	=\int_{0}^{r}f_{p}(\frac{t}{c})(1-\frac{t}{r})dt.
\end{equation}

For a fixed parameter $r$, the probability of collision increases as $c=\left \|v_1-v_2\right \|_{p}$ decreases. The family of hash functions is $(r_1, r_2, p_1, p_2)$-sensitive, $p_1 = p(1), p_2 = p(c), r_2 /r_1 = c$. Therefore, this family of locality-sensitive hash functions can be used to solve the approximate nearest neighbor problem.

In order to widen the gap between the collision probability between the points with short distance and the points with far distance after mapping, E2LSH uses $k$ position-sensitive functions together to build the function family:
\begin{equation}
	\mathcal{G}=\{g:S\longrightarrow U^k \}
\end{equation}
where $ \mathcal{G} $ represents the union of $k$ position-sensitive functions, and $g(v) = (h1 (v),..., h_{k}(v))$, then each data point $v \in \mathbb{R}^{d}$' dimension can be reduced via the function $ g(v) \in \mathcal{G} $ to obtain a $k$-dimensional vector $ \vec{a} = (a_1, a_2,..., a_k ) $. Then, E2LSH uses the main hash function $H_1$ and the secondary hash function $H_2$ to hash the vector after dimension reduction and establishes the hash table to store data points. The specific forms of $H_1$ and $H_2$ are as follows:

\begin{align}
	&H_1=((a_1*h_1+...a_k*h_k) \ mod \ C) \ mod \ T \\
    &H_2=(b_1*h+...b_k*h_k)\ mod\  C
\end{align}
where $a_i$ and $b_i$ are randomly selected integers, $T$ is the length of the hash table (generally set to the total number of data points $n$), and $C$ is a large prime number (can be set to $2^{32}-5$ on a 32-bit machine). Data points with the same primary hash value $H_1$ and secondary hash value $H_2$ will be stored in the same hash bucket to realize the clustering of data points.

For the query point $q$, E2LSH first uses the locality-sensitive hash function to obtain a set of hash values, then use $H_1$ to obtain its location in the hash table and then calculates its $H_2$ value, and obtain the same $H_2$ value of point $q$ by querying the linked list of the location point. Finally, to obtain a set of recovered points by querying $L$ tables and $K$ (or less than $K$) neighbor points by sorting the distances.

\section{The Framework of DNN-Based Learned Index}
\label{method}
Traditionally, we view index structure and machine learning algorithms as pretty different research branches. The index structure is constructed fixedly, but the machine learning algorithm is based on data training. However, both of them are positioning and searching for the space position. There is a potential connection between neural networks and indexes. A hash index can be regarded as a regression or a classification where the data is predicted based on the key, which is not fundamentally different from the neural network's. Inspired by the structure of the learned index, we propose the following groundbreaking work. This section will present our learned locality-sensitive hashing index framework in detail.

\subsection{The framework of DNN-based learned index}
\label{subsec:framework}
The ideal locality-sensitive hashing requires mapping and querying efficiently. Since the neural network with enough parameters has a robust fitting ability, using a deep neural network to simulate the hash function is meaningful. Empirically, an arbitrarily complex dataset fed into a well-trained model can always obtain the ideal mapping results.

The scheme of the proposed method (shown in Fig. \ref{fig:1}) can be divided into four stages: Input Stage, Autoencoder Stage, Neural Network Stage, and Hash index Stage. The supervised strategy is used to train the model in Neural Network Stage. When LLSH is trained, it can infer the input data to get the corresponding hash value. Taking image data as an example, each piece of data will go through the following four stages: 1) Feature extraction: where SIFT or GIST are generally used for feature extraction; 2) Dimensionality reduction: which refers to further dimensionality reduction of the extracted features by autoencoder; 3 ) Hash value generation: input the dimensionality-reduced feature vector into the neural network to generate corresponding hash value; 4) Hash index: perform the nearest neighbor search of the generated hash values. Since training the model in a supervised manner, it can guarantee that similar data will generate similar hash values. 

\subsubsection{Input Stage}
The input stage includes all kinds of data that require LSH to get mapping results in industrial or other scenarios, including various images, audio and text. Among them, some simple data, such as latitude and longitude data, can be directly input into the neural network. In contrast, other complex data need to be preprocessed (e.g., by feature extraction) before input into the LLSH, such as image data, audio data and et al. For example, the image data can use the GIST \cite{DBLP:journals/ijcv/OlivaT01} or SIFT \cite{DBLP:journals/ijcv/Lowe04}, and the audio data can use the MFCC \cite{DBLP:conf/iscas/LiuSK21} and the text data can use word2vector \cite{DBLP:conf/nips/MikolovSCCD13} to extract features, respectively.

\subsubsection{Autoencoder Stage}
Although the raw data has been extracted through the traditional feature extraction method, its correlation information needs to be expressed more adequately and the dimension of extracted feature is still too large, resulting in large amounts of parameters and further increasing computing consumption in the neural network part. So, LLSH first builds and trains an autoencoder model with a large amount of data to make it perform well and further reduce the extracted features dimensions regarding semantics. 

\subsubsection{Neural Network Stage}
The Neural Network Stage is the most critical part of the LLSH algorithm. In this stage, the neural network is composed of multiple DNN models, and the purpose is to encode the processed data. In this paper, we use $L$ neural networks to simulate $L$ locality-sensitive hashing function families, each of them outputs $k$ hash function values, the same as a traditional local-sensitive hash function at query time. In this way, we only need a set of neural networks that return the same result for similar data. 

Each neural network mentioned above acts as a family of hash functions, where the number of layers and neural nodes is determined according to the original hash structure. In the training process, we concat each neural network's output as the final output of the whole neural network stage to calculate the loss with the given label and further update the neural network's parameters. 
The training process will be finished soon because the parameters of each neural network are updated in a parallel way and do not affect each other. Besides, for NN search, we don't need each neural network's output exactly be the same.

\subsubsection{Hash Index Stage}
Finally, after the entire framework is well-trained, each neural network's output will be used as the hash index value and build the multiple hash tables. Empirically, the multiple hash tables can significantly reduce false-positive and false-negative rates \cite{DBLP:conf/compgeom/DatarIIM04}. In the querying, if the neural network outputs are the same for two input data, LLSH regards them as similar and maps them into the same storage address (bucket). For more convenience to find the index and decrease the computation when building a hash table, we build two extra hash functions, $H_1, H_2$, to transform the upper stage's output. The $H_1,H_2$ is shown below:
\begin{figure}[ht]
	\centering
	{\includegraphics[width=3.0in]{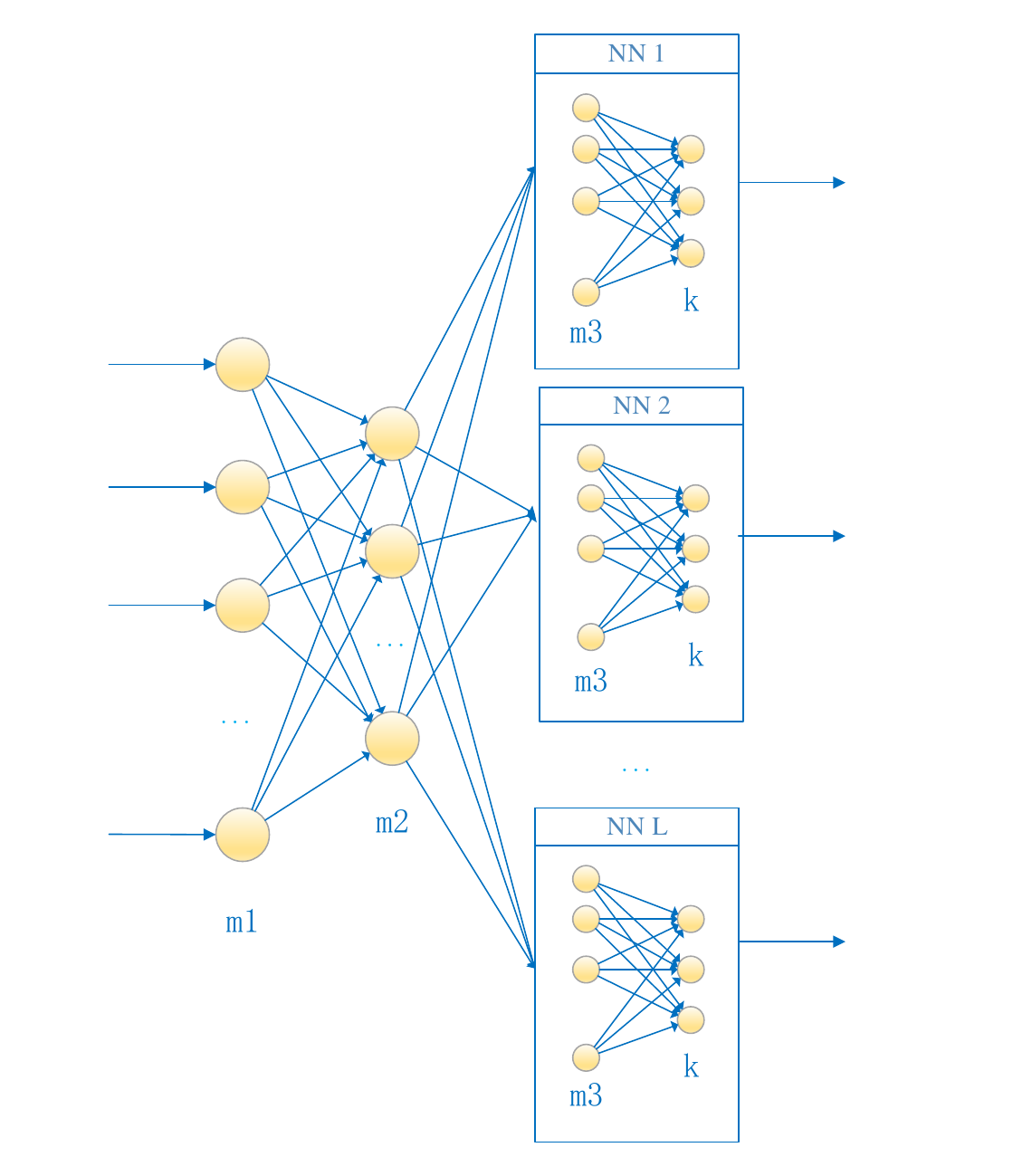}}
	\caption{The detail of autoencoder and neural network}
	\label{fig:3}
\end{figure}

\begin{equation}
	H_1(x_1,...,x_k)=((\sum_{i=1}^k{r_ix_i})\ mod\ C)\ mod\ T,
\end{equation}
\begin{equation}
	H_2(x_1,...,x_k)=(\sum_{i=1}^k{r_i'x_i})\ mod\ C),
\end{equation}
where $r_i,r_i'$ are random integers. $C=2^{32}-5$, is a large prime number. The $H_2$'s result is a data fingerprint, and the $H_1$'s result is the index of the hashtable in which the data fingerprint resides.

\begin{table*}[!htp]
	\centering
	
	%\textbf{Table 1}~~Improved table.
	\renewcommand\arraystretch{2}
	\caption{The datasets used in the experiments: four are randomly generated, and four are from the real world.}
	\resizebox{\textwidth}{20mm}{
		\begin{tabular}{cccccccccc}
			%     \specialrule[]{1}{2}
			%\centering
			\hline
			Dataset     & Type   & Dimension & Mean & Std  & Dataset     & Type     & Dimension & Mean  & Std   \\
			\hline
			Uniform     & Random & 100       & 0.5  & 0.29 & Tiny Images & GIST     & 384       & 0.11  & 0.07  \\
			\hline
			Normal      & Random & 100       & 0    & 1    & Ann SIFT    & SIFT     & 128       & 27.05 & 35.89 \\
			\hline
			Lognormal   & Random & 100       & 1.65 & 2.16 & Nytimes     & word2vec & 250       & 0     & 0.06  \\
			\hline
			Exponential & Random & 100       & 1    & 1    & Golve       & word2vec & 200       & 0     & 0.45  \\
			\hline
			
			%\caption{Datasets}
			%\bottomrule
		\end{tabular}
	\label{tab:dataset}
  
	}
\end{table*}

\begin{figure*}[htp]
	\centering	
    \includegraphics[width=0.9\textwidth]{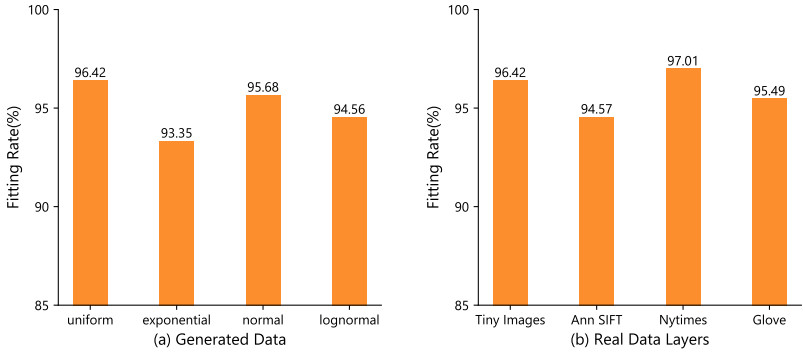}
	\caption{The fitting rate on different datasets.}
	\label{fig:4}
\end{figure*}

\subsection{Autoencoder and Neural Network Design}
In this subsection, we will introduce the autoencoder and neural network of the Sec. \ref{subsec:framework} in detail, including the architecture and parameter design. Fig. \ref{fig:3} shows the autoencoder and neural network detail.

To reduce the number of parameters in LLSH, we design a relatively small autoencoder that only includes the input layer, one hidden layer, and the output layer. In this paper, the autoencoder part is used as the feature extractor to reduce the data's dimension. To train this autoencoder more efficiently, we first pre-train it with large-scale data, and when faced with different datasets, we use transfer learning to fine-tune it again. When trained, the autoencoder can output a feature vector with a lower dimension.

Similar to the autoencoder, we use two fully connected (FC) layers to implement each small neural network unit. The whole model contains $N$ ($ N =1,2,..., L $) small units, named NN $L$ respectively. The first layer of each small unit contains $m3$ neurons and the last layer contains $k$ neurons. Each neural network's output contacts the final hash values of the whole model. Note that where the $N$, $m3$, and $k$ could adjust to keep good performance concerning the data size flexibly. 

Therefore, the number of LLSH's parameter is $p_1 = d*m_1+m_1*m_2+(m_2*m_3+m_3*k)*L$, while the traditional E2LSH algorithm is $p_2 = d*k*L$. In the actual implementation, we make $p_1<<p_2$ but without loss in query performance.

%放在B部分去
\subsection{Model training and prediction}
In this subsection, we will describe the train and prediction of the LLSH in detail. The first is to train the model to build the hash index well. When the model is well-trained, the second is to calculate the hash value of the query data by model prediction. Fig. \ref{fig:2} shows the overall framework of the first part. The specific steps are as follows:

\begin{itemize}
    \item Step 1: The feature extraction of different kinds of data such as images, audio, and texts extract features to obtain their corresponding feature vectors (v1), and then put the extracted feature vector into the autoencoder mentioned above to get more condensed vectors (v2) with lower dimension;
    
    \item Step 2: Input the vectors (v2) obtained by step 1 into the traditional E2LSH to obtain the $L*k$ hash values and concatenate them into a matrix as the label;
    
    \item Step 3: Train neural networks with the vectors (v2) and  their corresponding labels obtained in Step 2 until the model reaches convergence. 

    \item Step 4: Input the query item to the well-trained model for predicting the hash value.     
\end{itemize}

% Each small neural network in training consists of two fully-connected layers with randomly initialized parameters. In forward propagation, each small neural network's output serves as the hash value of the data. Although the structure of different small neural networks is similar, each neural network's initialization and backpropagation processes are independent. 

\textbf{Loss function:} The purpose of training neural networks is to make its output match the E2LSH output by iteratively updating the networks' parameters. And we expect the predicted results to be as close as possible to the hash value generated by the E2LSH. So, we chose the mean square error (MSE) loss as the objective function as follows:
\begin{equation}
	Loss = \frac{1}{N}\sum_{i=1}^{N}(y_i-\hat{y}_i)^2,
\end{equation}
where $y_i$ represents the neural network's output, $\hat{y_i}$ refers to label, and $ N $ is the total number of data of per output. We use Adam for optimizing and Relu for the activation in the training process.

% When the whole framework is well-trained, we can input query data into the model to calculate the hash value by forwarding. And then, we will obtain its fingerprint with the calculated hash value in the hash index stage. Finally, we will get the $k$ nearest neighbors from the database. 

\section{Experiment}
\label{experiment}
In this section, we will discuss the experiment details. All the experiments were conducted on a GPU server equipment with 128GB memory, two 2.1 GHz Intel(R) E5 processors, and two GTX1080Ti GPU cards with 11GB dedicated memory, and the operating system is CentOS 7. We use Python 3.6 and TensorFlow 1.13.1 to implement all code work. We repeat each experiment ten times and then use the median or average of the ten results as the final performance. 

\subsection{Setup}
\textbf{Datasets: } The dataset used in our experiments comes from two different types, synthetic data and real data, to adapt to data with different distributions in practical applications. Specifically, there are four synthetic datasets sampled from the distributions of uniform, exponential, normal and lognormal, respectively. The other four datasets include Tiny Images, Ann Sift, Nytimes and Glove, which involved images and word vectors. The details of the aforementioned datasets, which are detailed in Table \ref{tab:dataset}, contain different types, scales and dimensionality.

\textbf{Metrics: } In order to evaluate the simulating ability of the neural network-based LLSH in this work to the traditional E2LSH method, we use fitting accuracy as its evaluation metric, which refers to the correct rate of fitting E2LSH. It defines as follows:
\begin{equation}
	F_{rate}=\frac{M}{N}\times100\%,
\end{equation}
where the $M$ represents the same output numbers of neural networks and E2LSH, $N$ is the output dimension. A higher fitting accuracy indicates LLSH fits E2LSH more correctly. 

\textbf{Parameters: } For all of our experiments, we set the E2LSH parameters as $K=10, L=30$, and $r=4$ (the width of projection), and set $M=2$, $L=30$ and $K=10$ for the proposed LLSH.

\subsection{Ablation Study}
The suitable combination of parameters of $L, k, r$ significantly impacts the performance of traditional E2LSH. In our framework, however, the most critical parameters are $M, L, k$, where $M$ is the number of neural network layers. Therefore, in this subsection, we study how combining these parameters could boost LLSH.

\begin{figure*}[ht]
	\centering	
	\subfigure[]{
		\begin{minipage}[t]{0.5\linewidth}
			\centering
			\includegraphics[width=3.2in]{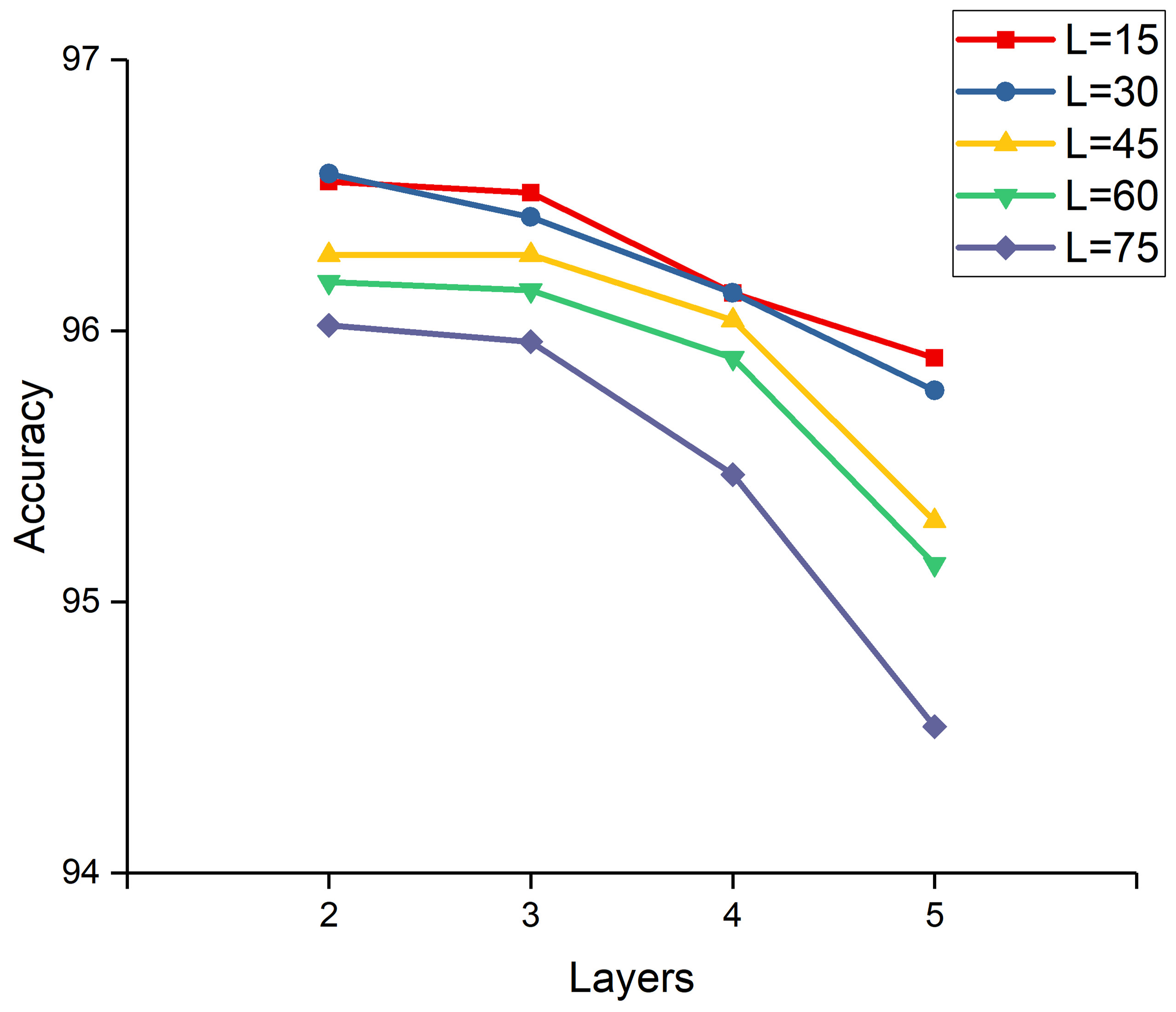}
			%\caption{fig1}
		\end{minipage}%
	}%
	\subfigure[]{
		\begin{minipage}[t]{0.5\linewidth}
			\centering
			\includegraphics[width=3.2in]{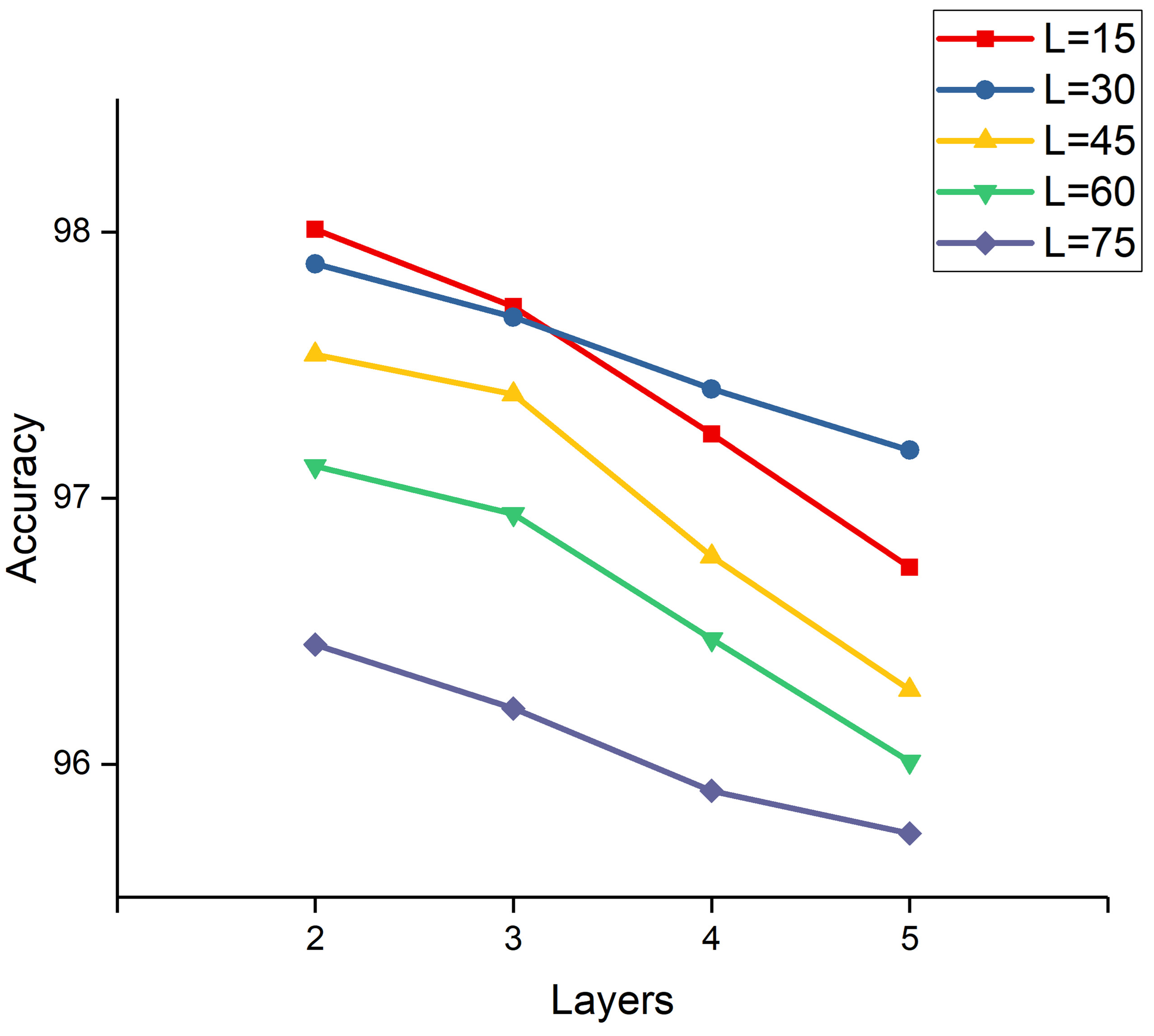}
			%\caption{fig2}
		\end{minipage}%
	}\\
	\subfigure[]{
		\begin{minipage}[t]{0.5\linewidth}
			\centering
			\includegraphics[width=3.2in]{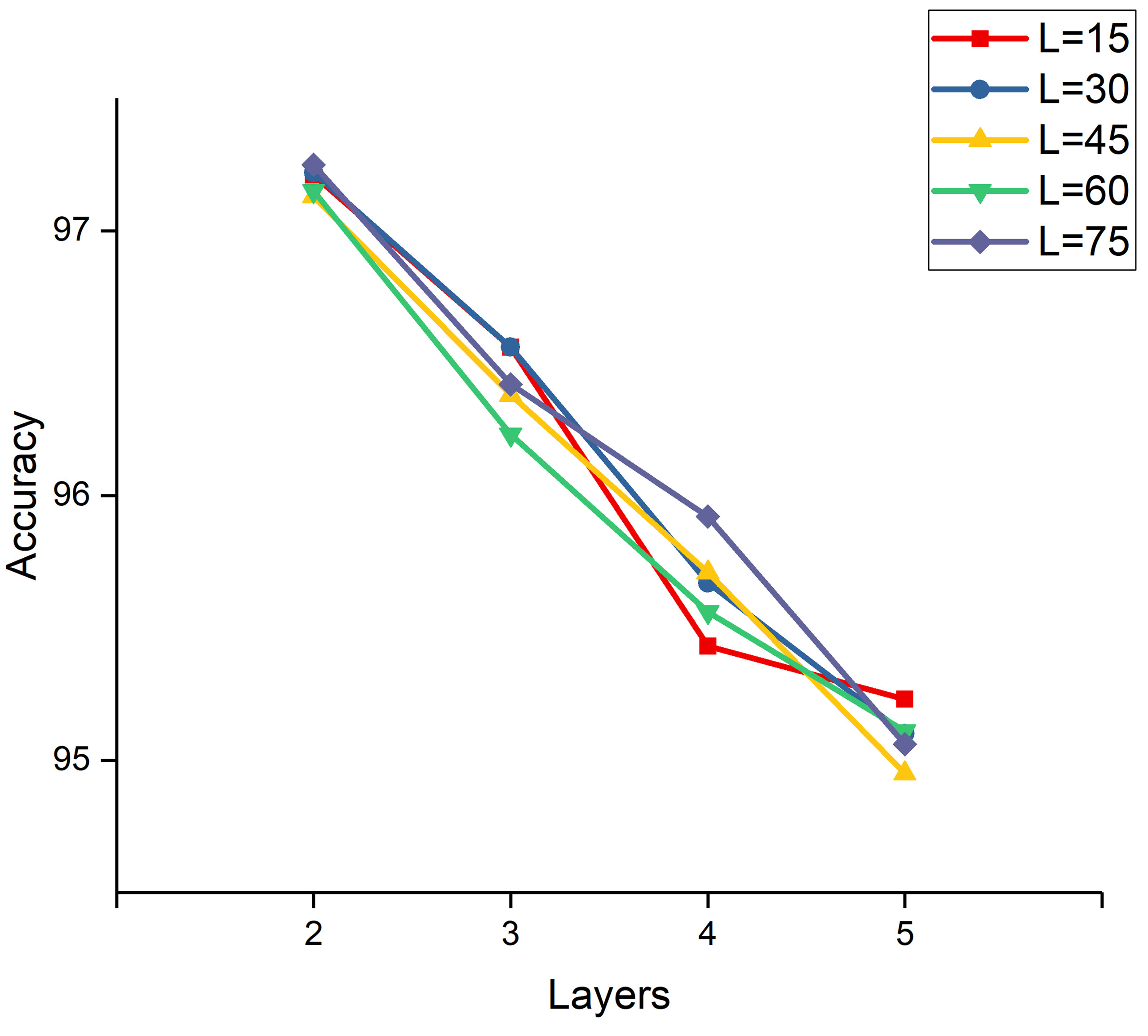}
			%\caption{fig2}
		\end{minipage}
	}%
	\subfigure[]{
		\begin{minipage}[t]{0.5\linewidth}
			\centering
			\includegraphics[width=3.2in]{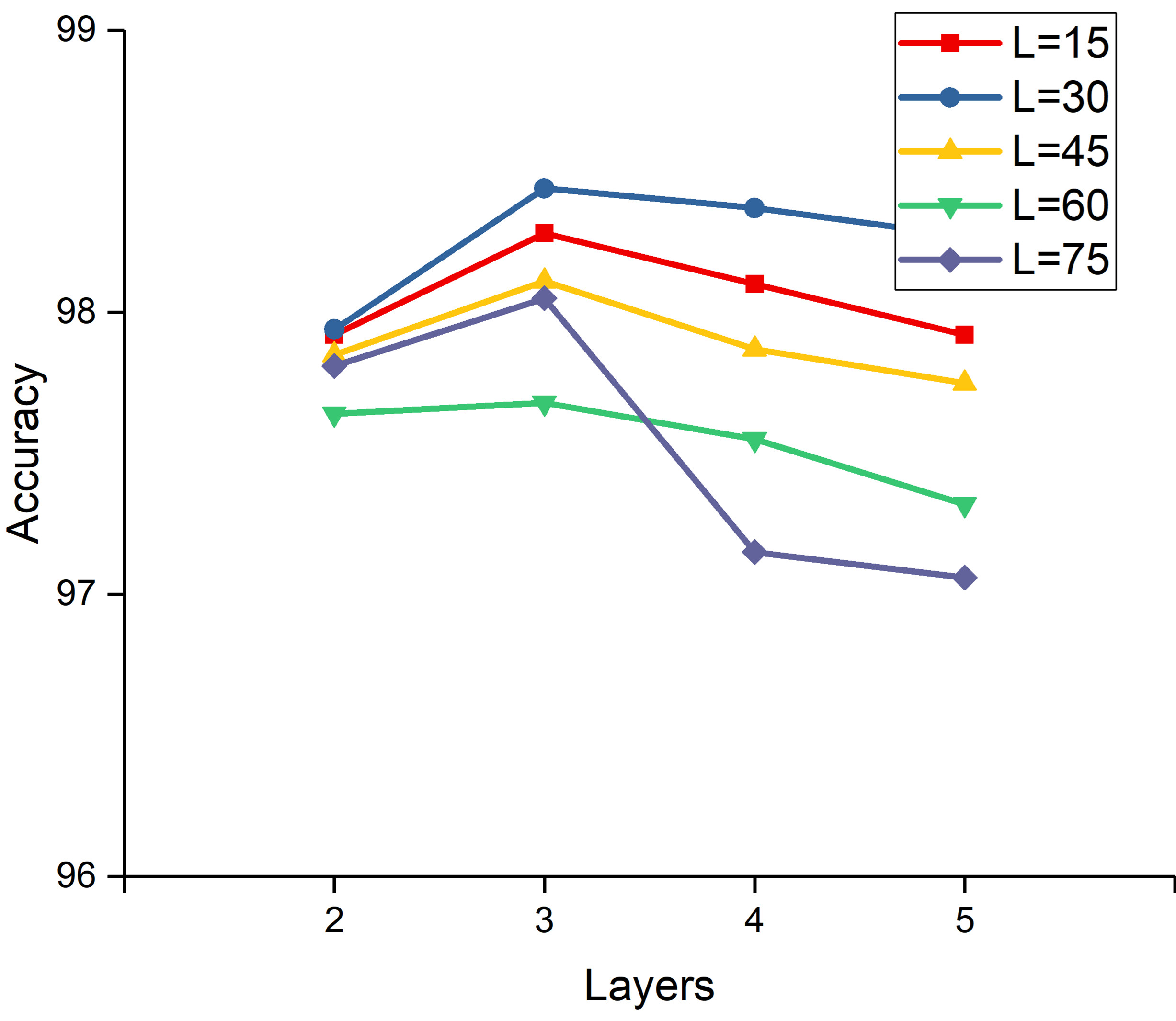}
			%\caption{fig2}
		\end{minipage}
	}
	\centering
	\caption{The fitting rate of the various number of neural networks $L$.}
	\label{fig:para_L}
\end{figure*}

\begin{figure*}[ht]
	\centering
	
	\subfigure[]{
		\begin{minipage}[t]{0.5\linewidth}
			\centering
			\includegraphics[width=3.2in]{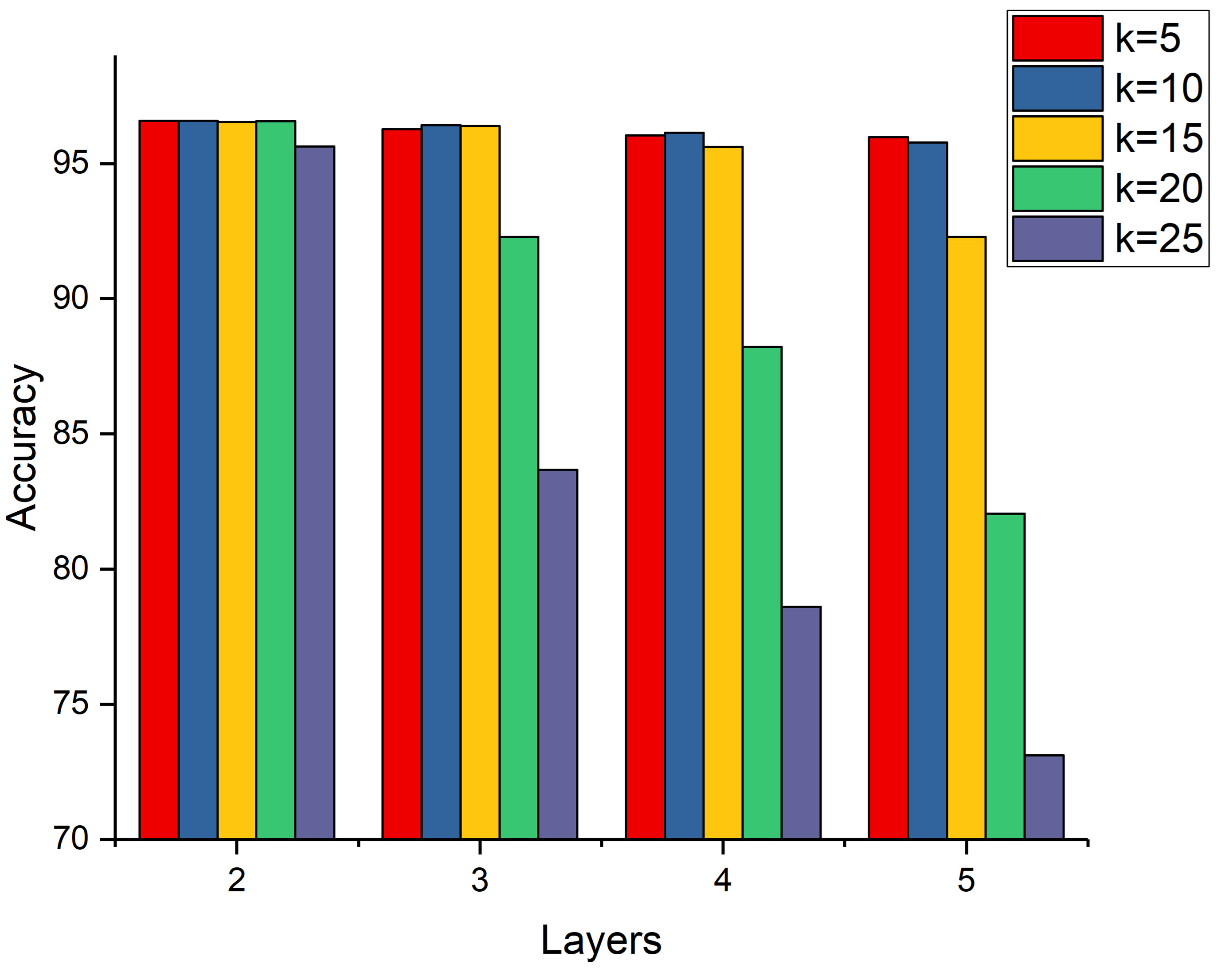}
			%\caption{fig1}
		\end{minipage}%
	}%
	\subfigure[]{
		\begin{minipage}[t]{0.5\linewidth}
			\centering
			\includegraphics[width=3.2in]{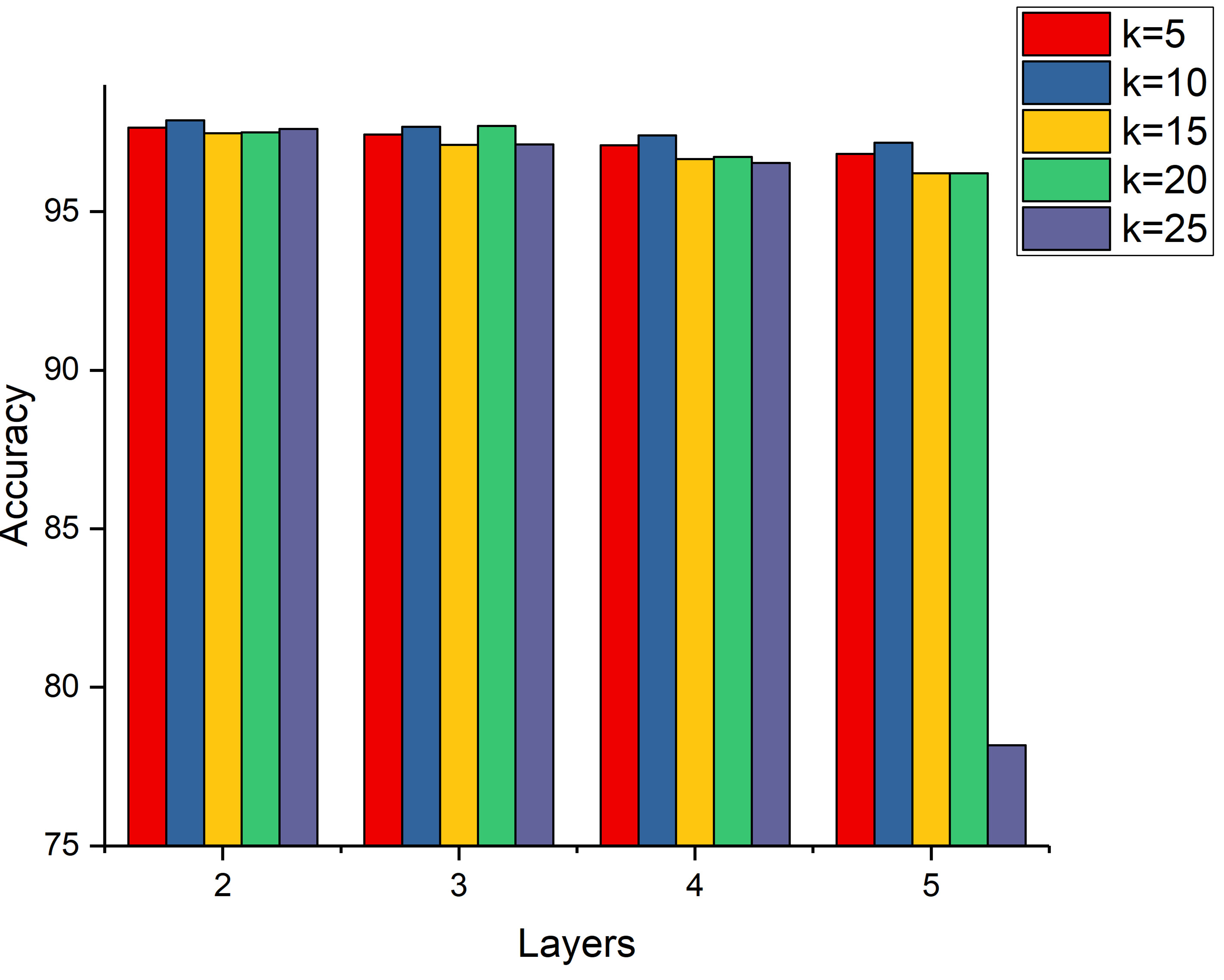}
			%\caption{fig2}
		\end{minipage}%
	}\\
	\subfigure[]{
		\begin{minipage}[t]{0.5\linewidth}
			\centering
			\includegraphics[width=3.2in]{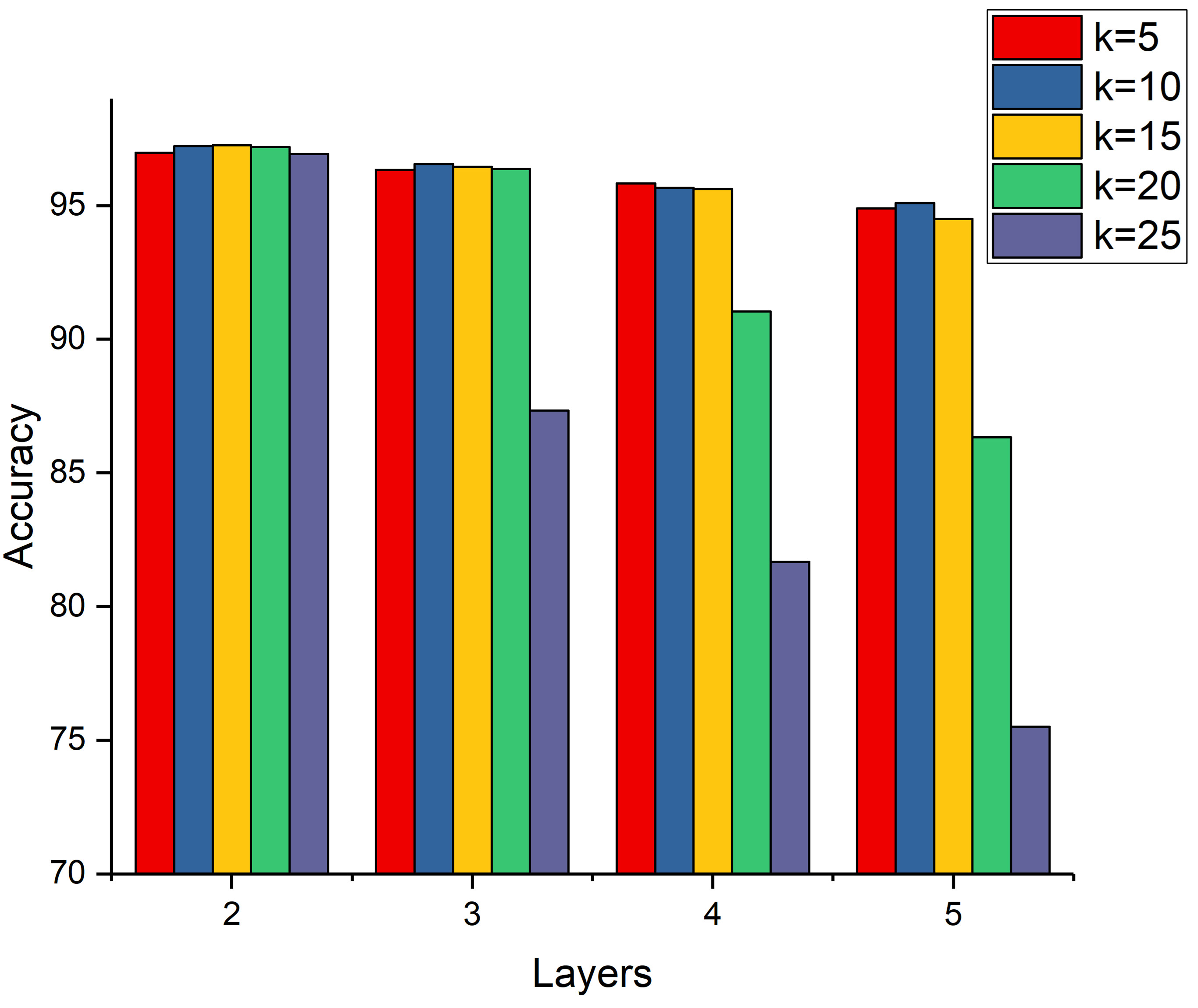}
			%\caption{fig2}
		\end{minipage}
	}%
	\subfigure[]{
		\begin{minipage}[t]{0.5\linewidth}
			\centering
			\includegraphics[width=3.2in]{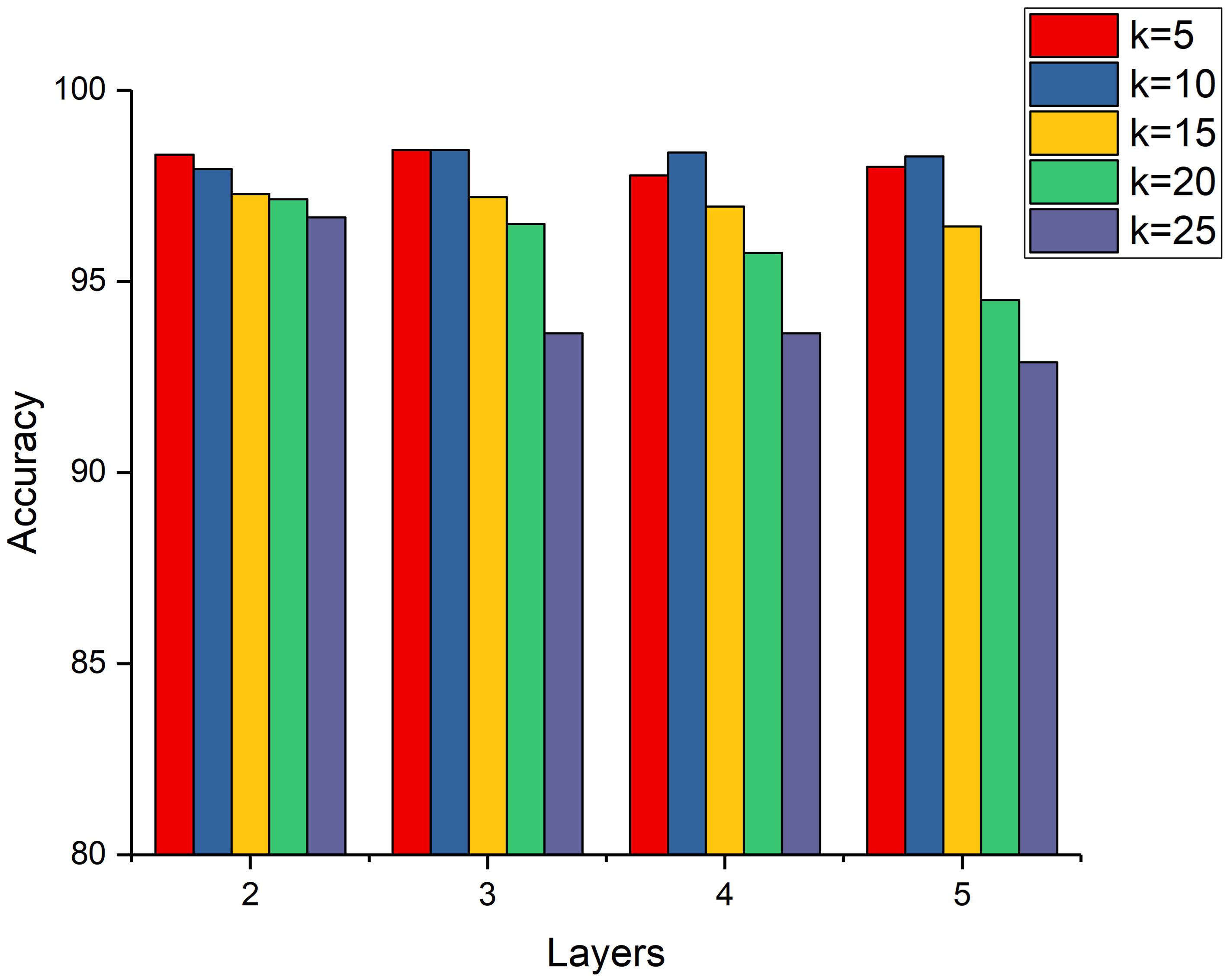}
			%\caption{fig2}
		\end{minipage}
	}
	\centering
	\caption{The fitting rate of the various number of nodes in the last layer $K$.}
	\label{fig:para_k}
\end{figure*}

In general, the more neural network layers mean the better the learning performance. However, our experiments show it is not exactly true for this work. The results in Fig. \ref{fig:para_L} and Fig. \ref{fig:para_k} illustrated the query accuracy of various $L$ on three random datasets drawn from uniform (a), normal (b) and lognormal distribution (c), and a real image dataset Tiny Images (d), respectively. The query accuracy decrease with the layers $M$ grows up, suggesting that a smaller $M$ shows a better query effect. The query accuracy reaches the top point when $M=2$ both in Fig. \ref{fig:para_L} and Fig. \ref{fig:para_k}. From Fig. \ref{fig:para_L}. We also observe that $L$ has a vital influence on query accuracy and $L=30$ leads to the highest query accuracy. Moreover, the results in Fig. \ref{fig:para_k} suggested that with the increase of $K$, the query accuracy decreases, and the best results can be obtained when $K=10$ in all cases.

Therefore, in the following experiments, we set the key parameters of the proposed LLSH as $M=2$, $L=30$ and $K=10$ to pursue optimal performance.

% In the previous subsection, we found suitable parameters $M=2, L=30, k=10$ according to the experimental results. Now, we will explore the advantages of LLSH with the right parameters. 

\begin{figure*}[ht]
	\centering
	\subfigure[]{
		\begin{minipage}[t]{0.5\linewidth}
			\centering
			\includegraphics[width=3.3in]{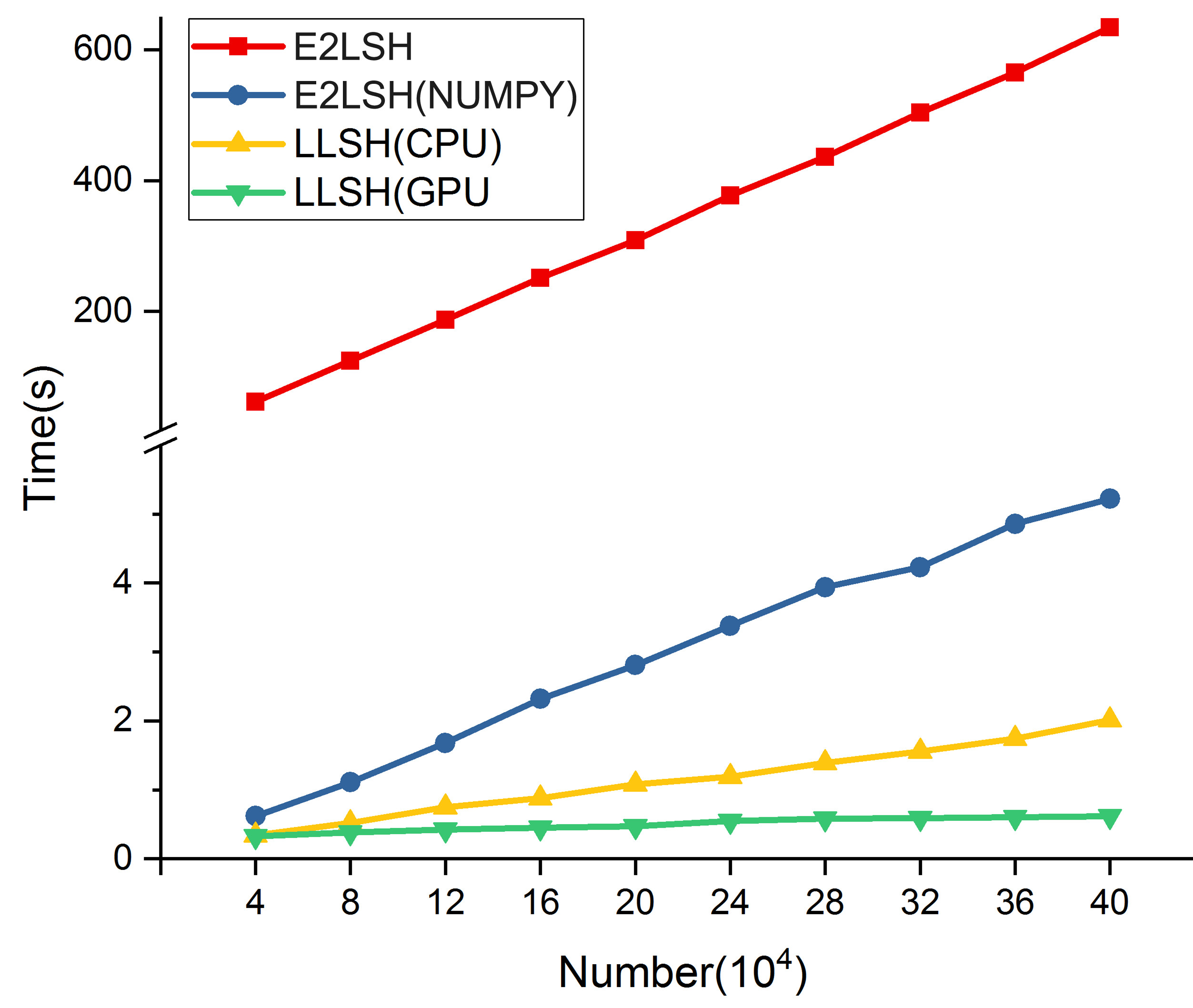}
			%\caption{fig1}
		\end{minipage}%
	}%
	\subfigure[]{
		\begin{minipage}[t]{0.5\linewidth}
			\centering
			\includegraphics[width=3.3in]{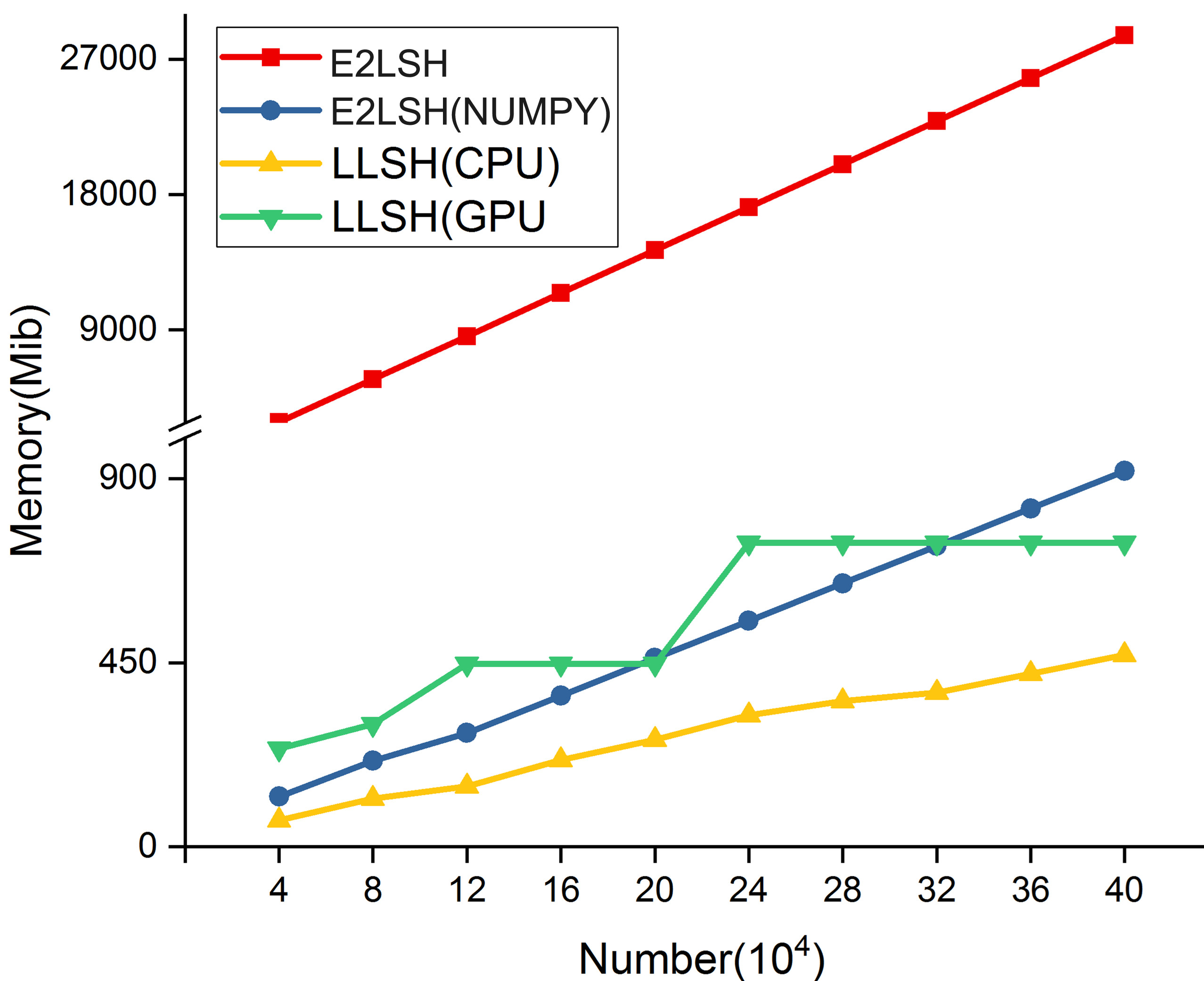}
			%\caption{fig2}
		\end{minipage}%
	}
	\centering
	\caption{Time \& Memory consumption vs. data with different magnitude.}
	\label{fig:workload_basic}
\end{figure*}

\subsection{Feasibility verification}
In practical applications, data often have intricate distributions. So, to verify the feasibility of the proposed LLSH to replace the traditional E2LSH, we carry conduct experiments on eight datasets from different distributions (which are detailed in Table. \ref{tab:dataset}) to verify whether LLSH can effectively fit the input-output mapping of E2LSH. The results are shown in Fig. \ref{fig:4}(a) and \ref{fig:4}(b) for the four synthetic datasets and four real-world datasets, respectively. 

The results in Fig. \ref{fig:4} show the neural network-based LLSH achieves excellent performance when fitting E2LSH. With randomly generated datasets, the fitting rates reach 96.42\%, 93.35\%, 95.68\% and 94.56\% on the uniform, exponential, normal and lognormal distribution, respectively. The result means that it is feasible for the LLSH to replace E2LSH with such a high fitting rate. Surprisingly, the experiments on four real datasets show that the fitting rates can grow up to 96.42\%, 94.57\%, 97.01\%, and 95.49\% on the Tiny Images, Ann SIFT, Nytimes, and Glove, respectively. The average fitting rate of LLSH on real datasets (95.87\%) is more significant than synthetic datasets (95.59\%), which shows that the LLSH framework can be deployed in physical scenarios.

\subsection{Evaluation of basic LLSH}
In this subsection, we design experiments in terms of memory and time consumption to evaluate LLSH's superiority in the process of hash value calculation. Here, we use the data with different magnitudes and dimensions and compare the performance of traditional E2LSH and its matrix-accelerated version (E2LSH(numpy)) and the LLSH running in different hardware (LLSH(CPU) and LLSH(GPU)).

\begin{figure*}[htp]
	\centering
	\subfigure[]{
		\begin{minipage}[t]{0.5\linewidth}
			\centering
			\includegraphics[width=3.2in]{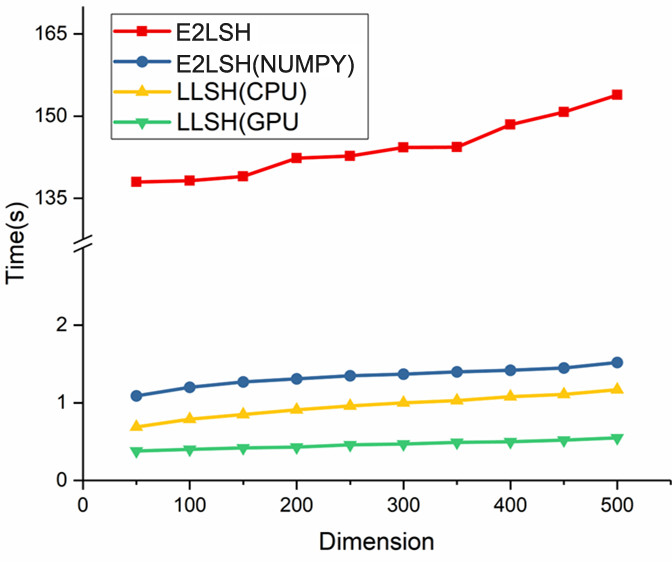}
			%\caption{fig1}
		\end{minipage}%
	}%
	\subfigure[]{
		\begin{minipage}[t]{0.5\linewidth}
			\centering
			\includegraphics[width=3.2in]{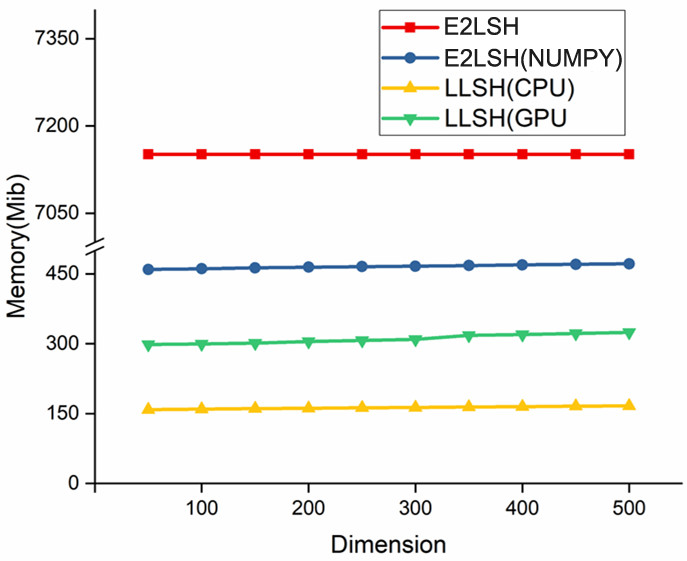}
			%\caption{fig2}
		\end{minipage}%
	}
	\centering
	\caption{Time \& Memory consuming vs. data with dimensions.}
	\label{fig:dimension_basic}
\end{figure*}

\begin{figure*}[ht]
	\centering
	\subfigure[]{
		\begin{minipage}[t]{0.5\linewidth}
			\centering
			\includegraphics[width=3.2in]{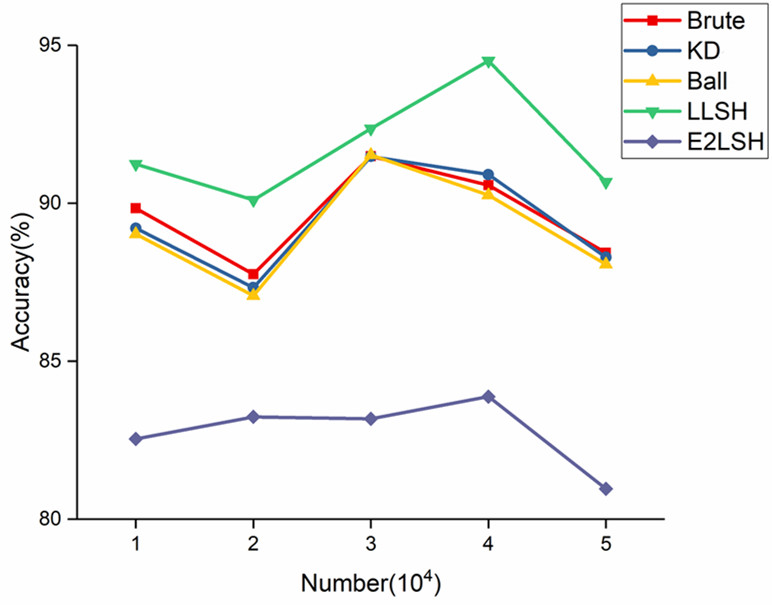}
			%\caption{fig1}
		\end{minipage}%
	}%
	\subfigure[]{
		\begin{minipage}[t]{0.5\linewidth}
			\centering
			\includegraphics[width=3.2in]{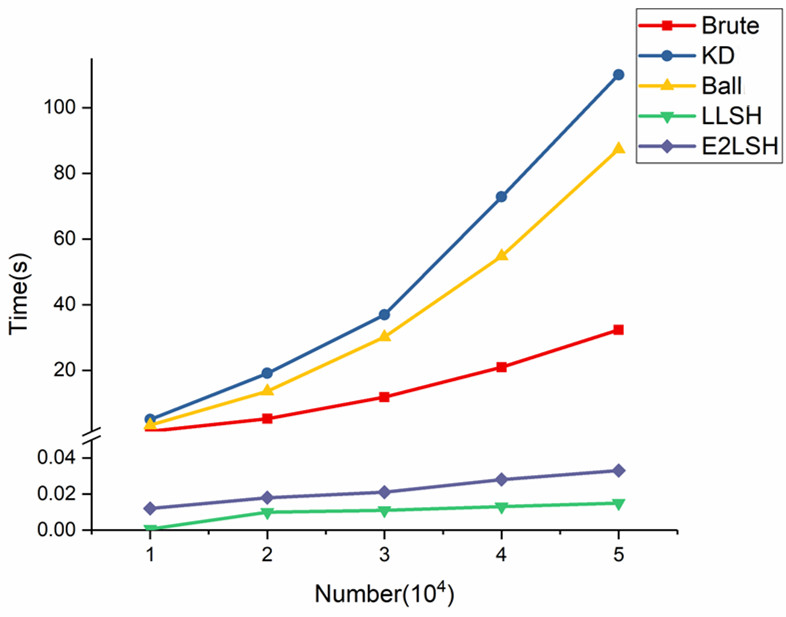}
			%\caption{fig2}
		\end{minipage}%
	}
	\centering
	\caption{Query accuracy \& Time consumption vs. data with different magnitude}

	\label{fig:ensemble_magnitude}
\end{figure*}

\begin{figure*}[ht]
	\centering
	\subfigure[]{
		\begin{minipage}[t]{0.5\linewidth}
			\centering
			\includegraphics[width=3.2in]{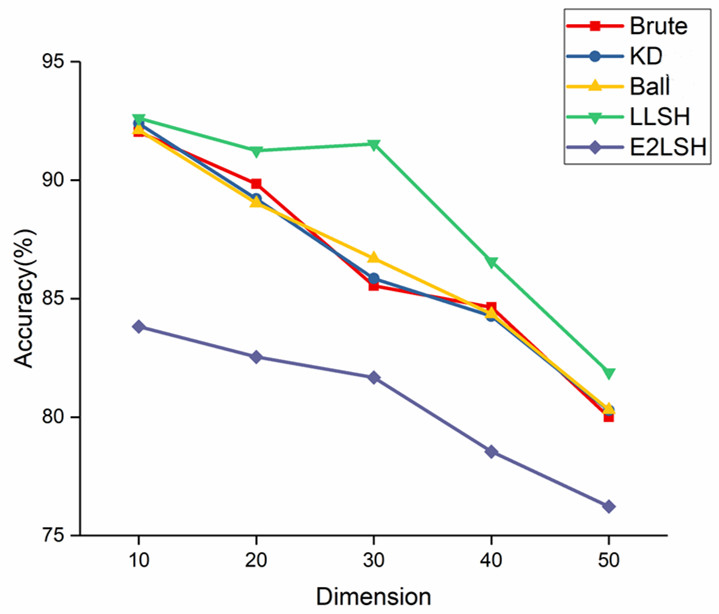}
			%\caption{fig1}
		\end{minipage}%
	}%
	\subfigure[]{
		\begin{minipage}[t]{0.5\linewidth}
			\centering
			\includegraphics[width=3.2in]{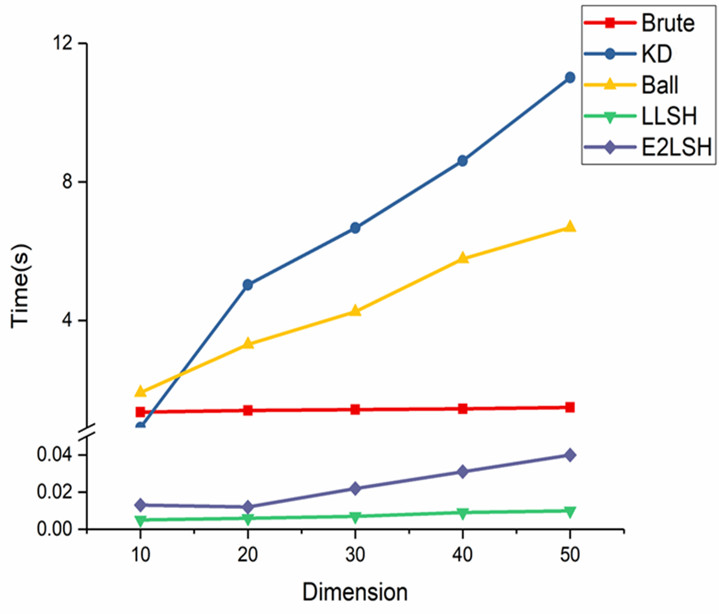}
			%\caption{fig2}
		\end{minipage}%
	}
	\centering
	\caption{Query accuracy \& Time consumption vs. data with different dimensions.}
	\label{fig:ensemble_accuracy}
\end{figure*}

For the data with different magnitudes, we draw data from a uniform distribution with a magnitude range from $1\times10^4$ to $4\times10^5$ as the validation dataset. As the results show in Fig. \ref{fig:workload_basic}(a), LLSH has an absolute superiority on time consumption, nearly 300 times faster than E2LSH in different data magnitudes. Moreover, as the data magnitude increases, the benefits continue to be improved. Compared with the matrix-accelerated E2LSH (E2LSH(numpy)), LLSH still has a nearly 50\% boost, and the advantages continue growing as the magnitude increases. Simultaneously, the LLSH can also be deployed on the GPU to fully use the advantages of new hardware, thus occupying great merit in large-scale data. As shown in Fig. \ref{fig:workload_basic}(b), LLSH has an overwhelming superiority in memory consumption in different data magnitudes, whereas the traditional E2LSH consumes memory about 40 more times than LLSH. The matrix-accelerated E2LSH's memory also consumes 1.7 times larger than the proposed LLSH. Compared to the CPU version, LLSH costs more memory on the GPU because part of the memory is consumed in exchange for a high computation speed.

For the dataset with different dimensions, we draw from a uniform distribution with dimensions ranging from 50 to 500 and keep the data magnitude as $1\times10^5$ to formulate the validation dataset. Fig. \ref{fig:dimension_basic} illustrates that LLSH is far beyond the traditional E2LSH. As shown in Fig. \ref{fig:dimension_basic}(a), different algorithms are insensitive to dimension, and the time consumption increases slowly with the dimension increase. The LLSH algorithm maintains significant advantages in any dimension, and the LLSH running deployed on GPUs shows greater advantages. Moreover, as shown in Fig. \ref{fig:dimension_basic}(b), the advantage of memory consumption is more obvious. The LSH consumes about 45 times less memory than the traditional E2LSH and about 2.8 times less than the matrix-accelerated E2LSH. 

According to the empirical results mentioned above, we found that LLSH has evident merits under various data magnitudes and dimensions. Benefiting from the fast reasoning ability of the neural network, the LLSH shows potential performance on time and memory consumption. This ability makes LLSH calculate faster than E2LSH when calculating the hash value. Moreover, LLSH's advantage is more pronounced on the new advanced computing device (GPU) with its parallel computing manner, its time consumption hardly increases as the data dimensions grow.

\subsection{Evaluation of ensemble-based LLSH}
LLSH aims to improve accuracy and reduce time and memory consumption. And ensemble learning can improve the accuracy of the model well. To make a step forward of the proposed LLSH, we introduce the ensemble strategy to LLSH, where different from the basic LLSH is the label for training is generated by multiple hash algorithms. We compare it with four traditional NN search methods, including Brute, KD-tree, Ball-tree and E2LSH.

In this experiment, the magnitude of the dataset is set from $1\times10^4$ to $5\times10^4$, and each dataset dimension is set to 20. The results in Fig. \ref{fig:ensemble_magnitude}(a) show that the ensemble-based LLSH has obtained higher accuracy than other baselines, even higher than the traditional tree-based algorithm by 2\% on average. While compared with the E2LSH, it is even more obvious and can achieve nearly 10\% higher. Regarding time consumption, as Fig. \ref{fig:ensemble_magnitude}(b) shows, the ensemble-based LLSH has extremely low time consumption; unlike the traditional tree algorithm, its time consumption will increase exponentially with the amount of data. The ensemble-based LLSH is nearly a hundred times faster than these tree-based algorithms, and the merits will be more evident with the larger data magnitude. Compared with the E2LSH, the improvement is nearly doubled. 

We also compared the ensemble-based LLSH and these four baselines' query accuracy and time consumption on the different data dimensions. Where the data dimension is set from 10 to 50 and the magnitude is set to $10^4$. As shown in Fig. \ref{fig:ensemble_accuracy}(a), the accuracy of different algorithms will decrease as the dimension increases, but the ensemble-based LLSH still has the best performance. In terms of time consumption, the results in Fig. \ref{fig:ensemble_accuracy}(b) show that the traditional tree algorithm's memory consumption will improve as the dimension increases, which is called the ``curse of dimension". So the tree-based algorithm is unsuitable for high-dimensional data. Besides, compared with E2LSH, the ensemble-based LLSH also shows its superiority in memory consumption. 

As discussed above, compared with the traditional hash algorithm, the ensemble-based LLSH can also improve accuracy and reduce time consumption. In addition, it has more comprehensive practical application value because it does not fall into the ``curse of dimension".

\section{Conclusions}
\label{conclusion}
In this paper, we investigated the LSH-based hash algorithms and the booming development of machine learning and high computing performance hardware. The traditional LSH-bash hash, however, is challenging to cope with the increasing dimensional and magnitude of massive data. To bridge this gap, we propose a novel learning-based hash framework, which uses multiple parallel neural networks to simulate the traditional hash functions to boost the hashing performance concerning time and memory consumption, and query accuracy. Extensive empirical results illustrated the feasibility of the proposed framework, and further showed its superiority in the effectiveness and efficiency of the NN search task with two implementations, i.e., the basic-based and the ensemble-based.

\backmatter

\section*{Compliance with Ethical Standards}

\begin{itemize}
\item \textbf{Funding } This work was partly supported by the National Natural Science Foundation of China under Grant 62162067 and the Yunnan Province Science Foundation under Grant No.202005AC160007, No. 202001B050076. And Open Foundation of Key Laboratory in Software Engineering of Yunnan Province under Grant No. 2020SE310. and Open Foundation of Engineering Research Center of Cyberspace under Grant No. KJAQ202112013.
\item \textbf{Competing interests} The authors declare that they have no competing of interests.
\item \textbf{Ethics approval} This article does not contain any studies with human participants performed by any of the authors.
\item \textbf{Informed consent} Not applicable.
\item \textbf{Consent to participate} Not applicable.
\item \textbf{Consent for publication} Not applicable.
\item \textbf{Data availability} The datasets used in this paper are available online publically.
\item \textbf{Code availability} Not applicable.
\item \textbf{Authors' contributions} All authors have equally contributed and all authors have read and agreed to the manuscript.
\end{itemize}

\bibliographystyle{sn-basic}
\bibliography{sn-bibliography}

\end{document}